\documentclass[10pt, conference]{IEEEtran}

\usepackage{amsmath}
\usepackage{amssymb}
\usepackage{booktabs}
\usepackage{cite}
\usepackage{graphicx}
\usepackage{microtype}
\usepackage{url}

\title{Shielded RL for Route-Charged Parity-Term Ordering in QEDA Phase Components}

\author{\IEEEauthorblockN{Owen Friedewald, Ali Shiri Sichani, and Chi-Ren Shyu}
\IEEEauthorblockA{Department of Electrical Engineering and Computer Science\\
University of Missouri, Columbia, MO, USA\\
Email: \{omfvq4, asp9f, shyuc\}@missouri.edu}}

\begin{document}
\maketitle

\begin{abstract}
Commuting phase terms in quantum electronic design automation (QEDA) placement circuits are logically invariant under reordering, yet their routed cost varies substantially after hardware mapping, since term order affects CNOT cancellation, interaction locality, and routing pressure. We cast parity/support phase-term ordering within a QEDA phase component as a shielded reinforcement-learning problem: a feasibility shield restricts each step to unemitted terms, so every trajectory is a valid permutation by construction, and an elite (cross-entropy-method) policy is trained against a route-charged proxy combining support-transition size and heavy-hex topology-distance features. We validate by direct Qiskit routing of logically equivalent circuits to a synthetic IBM-style heavy-hex map. On 36-term parity-walk components (50 term seeds $\times$ 2 transpiler seeds, statistics at the term-seed level), the per-instance learned ordering reduces mean routed CX to 336.0, a 5.7--12.2\% paired reduction over 2-opt and simulated-annealing search at equal or greater proxy budget and 22.3\% over the default construction order; routed-CX and routed-depth gains are significant after Bonferroni correction. Honest transfer audits show the proxy is predictive for the parity-walk component but not for extraction-heavy or token/permutation circuits, which require architecture-aware rewards, scoping the contribution accordingly.
\end{abstract}


\section{Introduction}

Quantum approximate optimization algorithms (QAOA) separate a cost phase from a mixer, making them a natural target for placement-like combinatorial problems~\cite{farhi2014qaoa}.  In this work, quantum electronic design automation (QEDA) denotes quantum-circuit constructions derived from EDA-style placement objectives.  A QEDA cost phase is often lowered into parity or support terms.  These terms commute, so their order does not change the logical diagonal operator.  The compiled circuit, however, is not order-invariant: term order affects CNOT cancellation, interaction locality, and routing pressure on constrained hardware.

This order sensitivity creates a constrained ordering problem within the phase component.  A decision is made sequentially, its cost depends on previous emitted supports, and invalid decisions must be excluded.  Shielded RL~\cite{alshiekh2018shielding} is well matched to this structure: the policy explores orderings, while a simple shield restricts actions to un-emitted terms and guarantees a legal permutation.  We focus on this parity-term ordering problem rather than end-to-end placement or ansatz design.

Our contributions are:
\begin{enumerate}
  \item We formulate QEDA parity/support ordering as a shielded RL MDP over a commuting phase component.
  \item We introduce a route-charged reward proxy combining logical support transition, synthetic heavy-hex support distance, and bridge distance.
  \item We show direct Qiskit routed-CX/depth reductions against the default construction order, greedy, hardware-distance, beam, and bounded local-search/insertion baselines.
  \item We characterize transfer boundaries showing that extraction-heavy and token/permutation circuits require architecture-aware reward models.
\end{enumerate}

\section{Method}

\subsection{Ordering MDP}

Let $\mathcal{T}=\{t_i\}_{i=1}^n$ be a multiset of parity/support terms.  Each term has a coefficient, a logical support set, and a fixed role in the phase separator.  We model ordering as a finite-horizon deterministic MDP $\langle \mathcal{S}, \mathcal{A}, P, r\rangle$.  A state $s=(O, m_{prev}, \iota_{prev})$ records the emitted prefix $O$, the support bitmask $m_{prev}$ of the previously emitted term, and its index $\iota_{prev}$ (with $m_{prev}=0$, $\iota_{prev}=\varnothing$ at $t=0$); the remaining set is $R(s)=\{1,\ldots,n\}\setminus O$.  An action $a\in\{1,\ldots,n\}$ appends a term, and the transition is deterministic: $O'=O\,\Vert\,a$, $m_{prev}'=\mathrm{mask}(a)$, $\iota_{prev}'=a$.  Episodes terminate when $R(s)=\emptyset$.

The shield is exact and enforced structurally, not by reward: the legal action set is restricted to the remaining terms,
\begin{equation}
\mathcal{A}(s) = R(s), \qquad \pi(a\mid s)=0 \;\; \forall\, a\notin R(s),
\label{eq:shield}
\end{equation}
and the policy is a categorical distribution indexed only over $R(s)$, so illegal actions are never assigned probability mass or sampled.  Because every action removes one remaining term, every episode returns a valid permutation of $\mathcal{T}$ regardless of $\pi$; correctness is a property of the action set, not of the learned weights.  Since every permutation of the commuting multiset is logically equivalent, this shield is a feasibility constraint over a commuting term set---reducing to sampling without replacement---rather than the safety shield of~\cite{alshiekh2018shielding}; our contribution is the route-charged proxy and its routed-cost reductions, not the masking mechanism itself.

The reward is the negative route-charged transition cost.  For candidate term $t_i$ after previous support $S_{prev}$, let $S_i$ be its logical support, $\phi(S)$ the fixed degree-heavy logical-to-physical map of a support, which ranks logical qubits by support frequency and coefficient magnitude and maps them to higher-degree, more central heavy-hex locations, $\mathrm{MST}$ the all-pairs heavy-hex minimum-spanning-tree distance over a mapped support, and $\mathrm{bridge}$ the closest heavy-hex distance between two mapped supports.  We use
\begin{align}
c(t_i\mid S_{prev}) &=
|S_i \triangle S_{prev}| + 0.35\,\mathrm{MST}(\phi(S_i)) \nonumber\\
&\quad + 0.15\,\mathrm{bridge}(\phi(S_{prev}),\phi(S_i)).
\label{eq:reward}
\end{align}
The proxy is used for learning and baseline search; routed CX and depth from direct transpilation are the evaluation metrics.  The episode return is the negative total transition cost of the completed ordering, $G(O)=-\sum_{t} c(O_t\mid S_{O_{t-1}})$, computed terminally rather than as a per-step reward.  The weights were fixed heuristically before direct-routing evaluation and were not tuned per routed instance; reward-weight tuning is left to future architecture-specific reward design.

\smallskip
\noindent\textbf{Algorithm 1 (shielded route-charged ordering).}
\emph{Input:} term multiset $\mathcal{T}$, proxy $c$, policy $\pi_\theta$.
For each training iteration: sample a batch of episodes, each starting from $R\leftarrow\mathcal{T}$, $O\leftarrow[]$ and repeatedly shielding actions to $R$, sampling $t\sim\pi_\theta(\cdot\mid O,R)$, appending $t$ to $O$, and removing $t$ from $R$ until $R=\emptyset$; assign each episode return $-\sum_j c(O_j,O_{j-1})$ and update $\theta$ via the elite (CEM-style) update~\eqref{eq:cvar}.  \emph{Output:} greedy decode of $\pi_\theta$, validated by direct routing.
\smallskip
\smallskip

\subsection{Policy and Training}

We use a stochastic linear policy over per-action features.  At each state, every legal candidate $a\in R(s)$ is scored by a feature vector $f(s,a)\in\mathbb{R}^{7}$ comprising the logical support-transition (Hamming) size $|S_a\triangle S_{prev}|$, the negative support overlap $-|S_a\cap S_{prev}|$, support cardinality, the heavy-hex MST emit cost $\mathrm{MST}(\phi(S_a))$, the bridge distance $\mathrm{bridge}(\phi(S_{prev}),\phi(S_a))$, the coefficient magnitude, and a bias term.  Note the policy feature space is richer than the three-term reward proxy of~\eqref{eq:reward}; the additional features inform ordering without altering the route-charged objective.  Logits are taken over legal actions only,
\begin{equation}
\pi_\theta(a\mid s) = \frac{\exp\!\big(-\theta^\top f(s,a)/\tau\big)}{\sum_{a'\in R(s)}\exp\!\big(-\theta^\top f(s,a')/\tau\big)},
\label{eq:policy}
\end{equation}
with temperature $\tau$; the negative sign assigns higher probability to lower learned linear cost.  For each 36-term instance, the policy is trained on that instance's shielded ordering MDP; the evaluation term seeds test the fixed ordering protocol rather than cross-instance generalization.

Training uses an elite (cross-entropy-method style) score-function (REINFORCE) update~\cite{sutton2000policy,williams1992reinforce,rubinstein2004cem}.  Each iteration samples a batch of $B$ complete orderings, computes terminal returns $G_i=G(O_i)$, and updates
\begin{equation}
\theta \leftarrow \theta + \frac{\eta}{B}\sum_{i=1}^{B}
(G_i - \bar{G})\,
\frac{\mathbf{1}\!\left[G_i \geq Q_{1-\alpha}\right]}{\alpha}\,
\nabla_\theta \log \pi_\theta(O_i),
\label{eq:cvar}
\end{equation}
where $\bar{G}$ is the full-batch mean baseline, $Q_{1-\alpha}$ is the $(1-\alpha)$ return quantile, and the indicator restricts the gradient to the top-$\alpha$ \emph{elite} tail of highest-return orderings (reweighted by $1/\alpha$).  This is an upper-tail, reward-seeking cross-entropy-method update, distinct from a Conditional Value-at-Risk objective (which averages the worst-return tail); the elite truncation makes~\eqref{eq:cvar} a biased estimator of the unmodified policy gradient, acceptable here since the goal is to surface the best orderings for greedy decoding.  We use $B=24$, $\alpha=0.25$, learning rate $\eta=0.02$, $\tau=1$, and $300$ iterations, with no value baseline, discount, entropy bonus, or gradient clipping.  The selected policy for direct validation is the greedy decode $\arg\max_{a\in R(s)}\pi_\theta(a\mid s)$.

All compared policies emit the same term multiset, supports, masks, and coefficients, differing only in order; since the terms commute, every ordering implements the same diagonal phase component.

\section{Experimental Setup}
\label{sec:setup}

The experimental object is a \emph{parity-walk component}: a circuit that emits a selected set of parity/support phase terms from the QEDA placement phase separator using a shared clean parity ancilla.  The component contains the ordered parity-walk network used to study order-sensitive routing; it does not include the full placement mixer, extraction scaffold, or complete circuit.  Each evaluation term seed deterministically samples 36 terms from the canonical phase-term pool generated by the flag-reuse phase construction; all compared policies route the same terms with the same coefficients and supports.

Circuits are routed with Qiskit~\cite{qiskit} to a synthetic IBM-style heavy-hex coupling map, \texttt{CouplingMap.from\_heavy\_hex(5)}, using basis gates \texttt{[rz, sx, x, cx]} and optimization level 3.  The direct-routing benchmark uses 50 evaluation term seeds and two transpiler seeds, yielding 50 $\times$ 2 = 100 routed rows per policy.  All policies are evaluated on identical term-seed/transpiler-seed pairs; no routing failures occurred in the main evaluation.  Statistical tests average the two transpiler seeds within each term seed, so the experimental unit is the term seed ($n=50$).

Baselines are defined as algorithms.  \emph{Default} is the deterministic construction order emitted by the existing parity-lowering pass before any route-aware, search-based, or learned reordering, serving as the unlearned construction-order baseline.  \emph{Greedy} repeatedly chooses the remaining term with minimum immediate route-proxy transition cost (666 proxy evaluations for 36 terms).  \emph{Beam} performs bounded beam search over complete orderings using the same proxy with beam width 8 ($\approx$9{,}792 proxy evaluations per instance).  \emph{Hardware-distance} greedily selects the next term minimizing heavy-hex emit plus bridge distance, using support-transition size only as a tie-breaker and omitting the full route-charged objective.  \emph{Insertion} places remaining terms into proxy-minimizing positions (8{,}400 evaluations).  We also evaluate 2-opt/local search and simulated annealing over stochastic swaps using 7{,}200 proxy evaluations, matching RL's 7{,}200 sampled orderings; the strongest such rows are warm-started from the Insertion and Beam orderings, so their total proxy budget (search plus warm start) exceeds RL's 7{,}200.  \emph{RL} is the shielded stochastic linear policy trained on route-charged returns and greedily decoded for evaluation.

\section{Results}
\label{sec:results}

Table~\ref{tab:main} gives the primary direct-routing result.  The learned route-charged policy achieves the lowest mean routed CX and depth among all tested policies, including 2-opt and simulated-annealing search baselines that received equal or greater proxy-evaluation budget.
\begin{table}[t]
\renewcommand{\arraystretch}{0.9}
\caption{Direct-routing benchmark on logically equivalent 36-term parity-walk components. Means over 50 term seeds $\times$ 2 transpiler seeds (100 rows/policy). Lower is better; proxy is the route-charged cost. $\pm$ are marginal SDs ignoring pairing; see Fig.~\ref{fig:bars} for paired CIs.}
\label{tab:main}
\centering
\scriptsize
\resizebox{\columnwidth}{!}{%
\begin{tabular}{llrrr}
\toprule
Policy & Family & Mean CX & Mean depth & Proxy \\
\midrule
RL elite full route-charged & RL & 336.0 $\pm$ 29.7 & 195.4 $\pm$ 25.8 & 145.02 \\
2-opt from Insertion & search & 357.1 $\pm$ 26.9 & 209.6 $\pm$ 26.3 & 149.09 \\
Route insertion & search & 361.4 $\pm$ 29.8 & 214.2 $\pm$ 29.2 & 150.38 \\
SA from Beam & search & 383.9 $\pm$ 34.9 & 227.2 $\pm$ 29.4 & 153.95 \\
Beam route-cost & baseline & 421.1 $\pm$ 42.6 & 245.0 $\pm$ 27.7 & 156.81 \\
Greedy route-cost & baseline & 421.5 $\pm$ 41.7 & 241.9 $\pm$ 25.6 & 159.59 \\
Default & baseline & 434.6 $\pm$ 39.7 & 298.5 $\pm$ 38.5 & 188.13 \\
Hardware-distance & baseline & 560.0 $\pm$ 59.1 & 320.6 $\pm$ 42.0 & 183.76 \\
\bottomrule
\end{tabular}
}
\end{table} Table~\ref{tab:winloss} summarizes paired wins after averaging the two transpiler seeds within each term seed.  The learned policy wins all routed-CX comparisons against Default, Hardware-distance, and Beam, and 49 of 50 against Greedy.  Against the strongest search rows, RL wins at least 39 of 50 routed-CX comparisons and at least 31 of 50 routed-depth comparisons.

\begin{figure}[t]
\centering
\includegraphics[width=0.85\columnwidth]{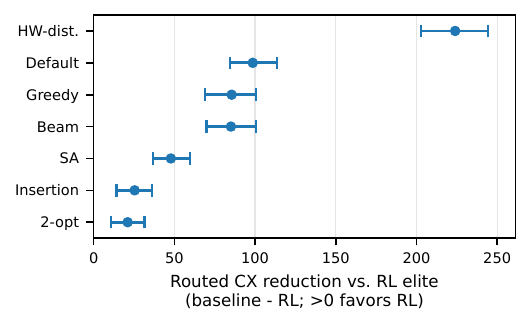}
\caption{Per-term-seed routed-CX reduction relative to the RL elite policy (baseline $-$ RL; positive favors RL), with Bonferroni-adjusted familywise 95\% bootstrap confidence intervals over the seven comparisons ($n=50$ term seeds). All intervals exclude zero, consistent with the term-seed-level Wilcoxon results in Section~\ref{sec:results}. Paired differences are shown rather than marginal standard deviations, which understate the paired design.}
\label{fig:bars}
\end{figure}

\begin{table}[t]
\renewcommand{\arraystretch}{0.9}
\caption{RL wins over references in term-seed-level paired comparisons (transpiler seeds averaged within term seed; same 50 term seeds as Table~\ref{tab:main}). Percentage reductions are averaged over paired term-seed cases, not from rounded means.}
\label{tab:winloss}
\centering
\scriptsize
\begin{tabular}{lrrrr}
\toprule
Reference & CX W/T/L & Depth W/T/L & CX red. & Depth red. \\
\midrule
Default & 50/0/0 & 50/0/0 & 22.33\% & 33.85\% \\
Greedy route-cost & 49/0/1 & 46/1/3 & 19.77\% & 18.49\% \\
Hardware-distance & 50/0/0 & 50/0/0 & 39.50\% & 38.01\% \\
Beam route-cost & 50/0/0 & 46/0/4 & 19.69\% & 19.25\% \\
2-opt from Insertion & 39/0/11 & 31/0/19 & 5.70\% & 5.55\% \\
SA from Beam & 47/0/3 & 39/0/11 & 12.16\% & 12.79\% \\
Route insertion & 39/1/10 & 34/0/16 & 6.75\% & 7.48\% \\
\bottomrule
\end{tabular}
\end{table}

The route proxy is predictive enough to guide search in this component setting over 800 routed policy-case rows spanning the eight Table~\ref{tab:main} policies.  Proxy versus routed CX has Pearson correlation 0.7481 and Spearman 0.7616; proxy versus routed depth has Pearson 0.7228 and Spearman 0.6847. These pooled correlations are descriptive across heterogeneous ordering families and should not be read as a within-family calibration. The proxy is most discriminating within proxy-aware ordering families: Default and Hardware-distance receive similar aggregate proxy costs but route very differently, so the proxy is not a universal circuit-quality estimator but a training/search signal whose usefulness depends on the ordering family.  This motivates the transfer audits in Section~\ref{sec:transfer}.A preliminary ablation (reported without the term-seed CIs of the main analysis) suggests the full route-charged formulation is best or near-best but does not establish that elite reweighting or every feature is individually necessary: mean-return and no-MST variants remain competitive, while removing bridge distance is weakest. We test at the term-seed level (n = 50; transpiler seeds averaged per term seed, Section III). One-sided Wilcoxon signed-rank tests on the paired differences keep all seven routed-CX comparisons significant after Bonferroni correction (largest corrected $p=5.5\times10^{-5}$). Routed-depth improvements also remain significant for all seven, the closest being 2-opt (corrected $p=2.0\times10^{-2}$). We report routed-CX reduction as primary and routed-depth reduction as secondary but statistically supported. A term-seed-level ablation with per-variant confidence intervals is left to future work.

\section{Robustness and Transfer Boundaries}
\label{sec:transfer}

We use transfer audits to identify where the route proxy remains predictive and where architecture-specific effects dominate. First, we inserted externally ordered terms into a fuller flag-reuse phase construction with extraction and unextraction.  The injection point was clean and logical equivalence passed.  The audit used three sampled 36-term phase seeds, one canonical full 124-term phase component, and one phase-plus-mixer smoke case.  In the sampled phase component, RL was best by proxy and mean depth, while the default construction order had the best mean routed CX.  In the full 124-term phase component, Insertion had the best routed CX and the default construction order had the best depth.  In this extraction-heavy setting, fixed scaffolding changes the routing objective and motivates extraction-aware rewards.

Second, we ported the controller to an explicit-EMPTY token/permutation architecture, a newer QEDA encoding in which real-cell and EMPTY tokens are permuted over site registers to preserve placement legality.  Initial validation on one instance was promising, but a broader audit across 27 instance/variant/transpiler configurations was stronger for the token default beam ordering, the beam-selected phase ordering used by the optimized token/permutation pipeline, and the default construction order overall.  RL still beat Greedy and Beam route-cost baselines, but proxy/direct correlations were weak or negative.  In token/permutation circuits, cancellation-aware ordering dominates the current route proxy, motivating token-specific rewards.

These transfer audits sharpen the contribution: shielded RL is effective for route-charged parity-component ordering, while extraction-heavy and token/permutation circuits require architecture-aware rewards.

\textbf{Threats to validity.}  The validation is component-level and
uses a synthetic heavy-hex coupling map.  Direct routing is
measured, but no noisy simulation or hardware execution is
reported.  The route proxy guides search in the parity-walk
component but does not transfer automatically to extraction-heavy
or optimized token/permutation circuits.  We compare routed
circuit quality under fixed ordering protocols, not optimizer
sample efficiency or wall time.  Both routed-CX and routed-depth
improvements survive Bonferroni correction at the term-seed level,
with 2-opt the closest depth comparison.  We therefore claim
ordering improvement for the studied component, not end-to-end
placement improvement. Code and logs are available at https://github.com/owenfriedewald/shielded-qrl.

\section{Related Work and Discussion}

Several lines of prior work apply RL to quantum compilation.
Pozzi et al.~\cite{pozzi2022rl} use deep Q-learning to route
quantum circuits by learning when and where to insert SWAP gates,
and Murali et al.~\cite{murali2019noise} extend heuristic mapping
to noise-adaptive settings; both address the physical placement
and SWAP-insertion subproblem, where the policy must learn not
to generate invalid hardware operations.
F\"osel et al.~\cite{fosel2021rl} train a deep convolutional
agent to discover gate-level rewrite rules for circuit
optimization, with validity maintained by restricting rewrites
to a known-legal rule set.
Quetschlich et al.~\cite{quetschlich2023rl} and Mills et
al.~\cite{mills2026passes} use RL to select and compose compiler
passes, treating pass selection rather than term ordering as the
decision variable.
Tang et al.~\cite{tang2024alpharouter} combine MCTS with RL for
routing.
Kremer et al.~\cite{kremer2024rl} integrate RL into Qiskit
transpiling, synthesizing Clifford and Permutation circuits up
to 65 qubits; their Permutation family is structurally related
to the token/permutation architecture in our transfer audits,
where cancellation-aware ordering not captured by our proxy
dominates.

In each of these lines, correctness is maintained by restricting
the action space to valid operations, by reward shaping, or by
post-hoc validation.  The present work differs in both problem
and guarantee: rather than mapping or routing, we order terms
inside a commuting phase component where every permutation is
logically equivalent by construction, so the
shield~\cite{alshiekh2018shielding} excludes invalid actions
entirely and correctness is structural, not learned.  This also
distinguishes our ordering layer from Li et
al.~\cite{li2019sabre}, whose SABRE heuristic addresses qubit
mapping and SWAP insertion at the gate level.
Our contribution is an ordering layer for
exactness-preserving decisions within an already-placed
QEDA phase component, leaving physical mapping to the downstream
Qiskit transpiler~\cite{qiskit}.

Recent work highlights the gate-count/fidelity gap our
proxy-based evaluation also exposes: Tomar et
al.~\cite{tomar2026calibration} show calibration-aware RL routing
on IBM Heron improving exact fidelity over SABRE while accepting
higher routed two-qubit counts, the trade-off our Threats to
Validity section flags for future work.  The route-charged proxy
and its transfer limits thus connect to active interest in
hardware-aware reward design, motivating the architecture-aware
extensions of Section~\ref{sec:transfer}.

\section{Conclusion}

Shielded RL can learn route-aware orderings of parity/support terms that preserve logical validity while reducing routed cost, achieving the lowest mean routed CX and depth in the parity-walk component against all tested baselines. Transfer audits show that extraction-heavy and token/permutation circuits require architecture-aware rewards. Future work should model cancellation and extraction structure directly, then revisit full-circuit, noisy, and hardware validation once routed-resource gains transfer.

\section*{Acknowledgment}
The computation for this work was performed on the University of Missouri's Quantum Innovation Center, in partnership with IBM Quantum and facilitated by Research Support Services at the University of Missouri, Columbia, MO. DOI: 10.32469/10355/107781.

\footnotesize
\bibliographystyle{IEEEtran}
\bibliography{references}

@article{farhi2014qaoa,
  author = {Farhi, Edward and Goldstone, Jeffrey and Gutmann, Sam},
  title = {A Quantum Approximate Optimization Algorithm},
  journal = {arXiv preprint arXiv:1411.4028},
  year = {2014}
}

@book{sutton2000policy,
  author = {Sutton, Richard S. and Barto, Andrew G.},
  title = {Reinforcement Learning: An Introduction},
  publisher = {MIT Press},
  edition = {2},
  year = {2018}
}

@inproceedings{li2019sabre,
  author = {Li, Gushu and Ding, Yufei and Xie, Yuan},
  title = {{Tackling the Qubit Mapping Problem for NISQ-Era Quantum Devices}},
  booktitle = {Proceedings of the Twenty-Fourth International Conference on Architectural Support for Programming Languages and Operating Systems},
  pages = {1001--1014},
  year = {2019},
  doi = {10.1145/3297858.3304023}
}

@inproceedings{murali2019noise,
  author = {Murali, Prakash and others},
  title = {Noise-Adaptive Compiler Mappings for Noisy Intermediate-Scale Quantum Computers},
  booktitle = {Proceedings of the Twenty-Fourth International Conference on Architectural Support for Programming Languages and Operating Systems},
  pages = {1015--1029},
  year = {2019},
  doi = {10.1145/3297858.3304075}
}

@article{williams1992reinforce,
  author = {Williams, Ronald J.},
  title = {Simple Statistical Gradient-Following Algorithms for Connectionist Reinforcement Learning},
  journal = {Machine Learning},
  volume = {8},
  pages = {229--256},
  year = {1992},
  doi = {10.1007/BF00992696}
}

@inproceedings{alshiekh2018shielding,
  author = {Alshiekh, Mohammed and others},
  title = {Safe Reinforcement Learning via Shielding},
  booktitle = {Proceedings of the AAAI Conference on Artificial Intelligence},
  year = {2018}
}

@misc{qiskit,
      title={Quantum computing with {Q}iskit},
      author={Javadi-Abhari, Ali and Treinish, Matthew and Krsulich, Kevin and Wood, Christopher J. and Lishman, Jake and Gacon, Julien and Martiel, Simon and Nation, Paul D. and Bishop, Lev S. and Cross, Andrew W. and Johnson, Blake R. and Gambetta, Jay M.},
      year={2024},
      doi={10.48550/arXiv.2405.08810},
      eprint={2405.08810},
      archivePrefix={arXiv},
      primaryClass={quant-ph}
}

@article{pozzi2022rl,
  author  = {Matteo G. Pozzi and others},
  title   = {Using Reinforcement Learning to Perform Qubit
             Routing in Quantum Compilers},
  journal = {ACM Transactions on Quantum Computing},
  volume  = {3},
  number  = {2},
  pages   = {1--25},
  year    = {2022}
}

@article{fosel2021rl,
  author  = {Thomas F{\"o}sel and others},
  title   = {Quantum Circuit Optimization with Deep
             Reinforcement Learning},
  journal = {arXiv preprint arXiv:2103.07585},
  year    = {2021}
}

@inproceedings{quetschlich2023rl,
  author    = {Nils Quetschlich and Lukas Burgholzer
               and Robert Wille},
  title     = {Compiler Optimization for Quantum Computing
               Using Reinforcement Learning},
  booktitle = {Proceedings of the 60th ACM/IEEE Design
               Automation Conference (DAC)},
  pages     = {1--6},
  year      = {2023}
}

@article{tang2024alpharouter,
  author  = {Wei Tang and others},
  title   = {{AlphaRouter}: Quantum Circuit Routing with
             Reinforcement Learning and Tree Search},
  journal = {arXiv preprint arXiv:2410.05115},
  year    = {2024}
}

@article{kremer2024rl,
  author  = {David Kremer and others},
  title   = {Practical and Efficient Quantum Circuit Synthesis
             and Transpiling with Reinforcement Learning},
  journal = {arXiv preprint arXiv:2405.13196},
  year    = {2024}
}

@article{mills2026passes,
  author  = {Daniel Mills and others},
  title   = {Reinforcement Learning for Adaptive Composition
             of Quantum Circuit Optimisation Passes},
  journal = {arXiv preprint arXiv:2601.21629},
  year    = {2026}
}

@article{tomar2026calibration,
  author  = {Yash Tomar and others},
  title   = {Graph Reinforcement Learning for
             Calibration-Aware Quantum Circuit Routing},
  journal = {arXiv preprint arXiv:2606.12816},
  year    = {2026}
}

@book{rubinstein2004cem, author={Rubinstein, R. Y. and Kroese, D. P.}, title={The Cross-Entropy Method: A Unified Approach to Combinatorial Optimization, Monte-Carlo Simulation and Machine Learning}, publisher={Springer}, year={2004}
}

\end{document}